\documentclass[WCMS,STIX2COL]{WileyNJDv5}

\usepackage{makecell} 
\usepackage{multirow}
\usepackage{booktabs}
\usepackage{graphicx}
\usepackage{caption}
\usepackage{placeins}

\articletype{Research Article}%

\received{Date Month Year}
\revised{Date Month Year}
\accepted{Date Month Year}
\startpage{1}

\begin{document}

\title{Cavity Formation in Silica-Filled Rubber Compounds Observed During Deformation by Ultra Small-Angle X-Ray Scattering}

\author[1,3]{Ilya Yakovlev}

\author[1]{Michael Sztucki}

\author[2]{Frank Fleck}

\author[2]{Hossein Ali Karimi-Varzaneh}

\author[2,3]{Jorge Lacayo-Pineda}

\author[2]{Christoph Vatterott}

\author[3,4]{Ulrich Giese}

\authormark{YAKOVLEV \textsc{et al.}}
\titlemark{Cavity Formation in silica-Filled Rubber Compounds Observed During Deformation by Ultra Small-Angle X-Ray Scattering}

\address[1]{\orgname{European Synchrotron Radiation Facility}, \orgaddress{\city{Grenoble}, \country{France}}, \email{sztucki@esrf.fr}}

\address[2]{\orgname{Continental Reifen Deutschland GmbH}, \orgaddress{\city{Hanover}, \country{Germany}}, \email{ali.karimi@conti.de, frank.fleck@conti.de, christoph.vatterott@conti.de}}

\address[3]{\orgdiv{Faculty of Natural Sciences}, \orgname{Gottfried Wilhelm Leibniz University Hannover}, \orgaddress{\city{Hanover}, \country{Germany}},\email{jorge.lacayo-pineda@conti.de}}

\address[4]{\orgname{Deutsches Institut für Kautschuktechnologie e. V.}, \orgaddress{\city{Hanover}, \country{Germany}},\email{ulrich.giese@dikautschuk.de}}

\corres{Corresponding author Ilya Yakovlev, European Synchrotron Radiation Facility, Avenue des Martyrs 71, Grenoble, 38043, France \email{yakovlev.iliya@gmail.com}, ORCID: 0000-0003-1730-6113}

\fundingInfo{H2020 Marie Skłodowska-Curie Actions grant number 847439 468 to the InnovaXN project.}

\abstract[Abstract]{When silica-filled rubber compounds are deformed, structural modifications in the material’s bulk lead to irreversible damage, the most significant of which is cavitation appearing within the interfaces of interconnected polymer and filler networks. This work introduces a new method to analyze cavitation in industrial-grade rubbers based on Ultra Small-Angle X-ray Scattering. This method employs a specially designed multi-sample stretching device for high-throughput measurements with statistical relevance. The proposed data reduction approach allows for early detection and quantification of cavitation while providing at the same time information on the hierarchical filler structures at length scales ranging from the primary particle size to large silica agglomerates over four orders of magnitude. To validate the method, the scattering of SSBR rubber compounds filled with highly dispersible silica at different ratios was measured under quasi-static strain. The strain was applied in incremental steps up to a maximum achievable elongation or breakage of the sample. From the measurements performed in multiple repetitions, it was found that the minimum strain necessary for cavity formation and the size evolution of the cavities with increasing strain are comparable between these samples. The sample with the highest polymer content showed the lowest rate of cavity formation and higher durability of silica structures. The structural stability of the compounds was determined by the evolution of the filler hierarchical structures, obtained by fitting data across the available strain range.}

\keywords{Ultra Small-Angle X-Ray Scattering (USAXS), Filled Rubber Compounds, Cavitation, \textit{In Situ} Straining}

\maketitle

\renewcommand\thefootnote{}
\footnotetext{\textbf{Abbreviations:} SAXS, Small Angle X-ray Scattering; USAXS, Ultra Small Angle X-ray Scattering; $\mu$CT, Micro-Computed X-ray tomography; TLM, Transmitted Light Microscopy; TEM, Transmission Electron Microscopy; SEM, Scanning Electron Microscopy; AFM, Atomic Force Microscopy; ESRF, European Synchrotron Radiation Facility; SBR, Styrene Butadiene Rubber; HD, Highly Dispersible; TDAE, Treated Distillate Aromatic Extract; DIN, German Institute for Standardization; ISO, International Organization for Standardization; M300, stress at 300 \% strain; USF, Unified Scattering Function}

\renewcommand\thefootnote{\fnsymbol{footnote}}
\setcounter{footnote}{1}

\section{Introduction}\label{intro}

\begin{figure*}[!b]
\centerline{\includegraphics[width=0.85\textwidth]{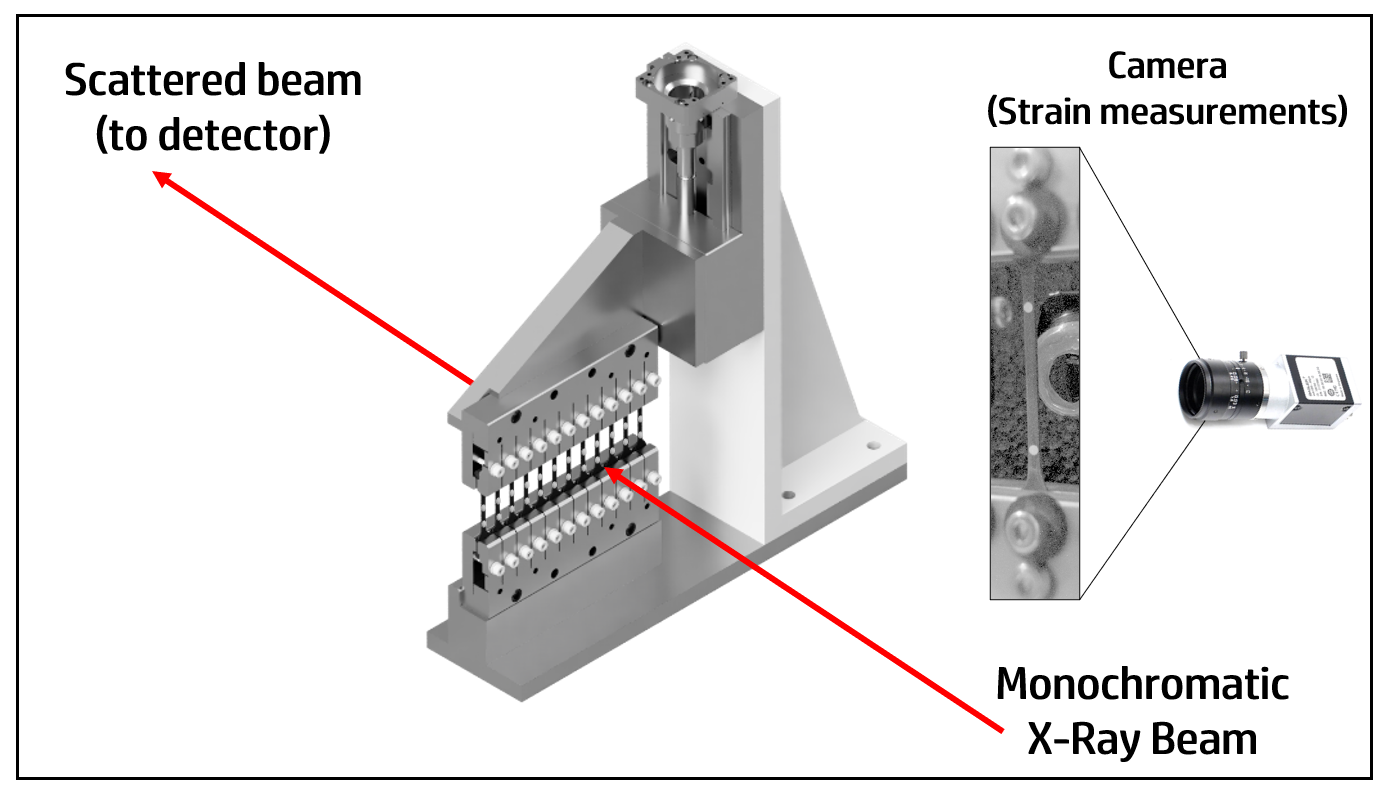}}
\caption{Scheme of the set-up used for the experiments: a specially designed multi-sample stretching device was positioned on the 3-axis sample table, allowing consecutive measurement of up to 15 samples; a monochrome camera was installed in front of the device to register the position of reflecting dots for strain calibration. \label{setup}}
\end{figure*}

\begin{figure*}[!t]
\centerline{\includegraphics[width=0.65\textwidth]{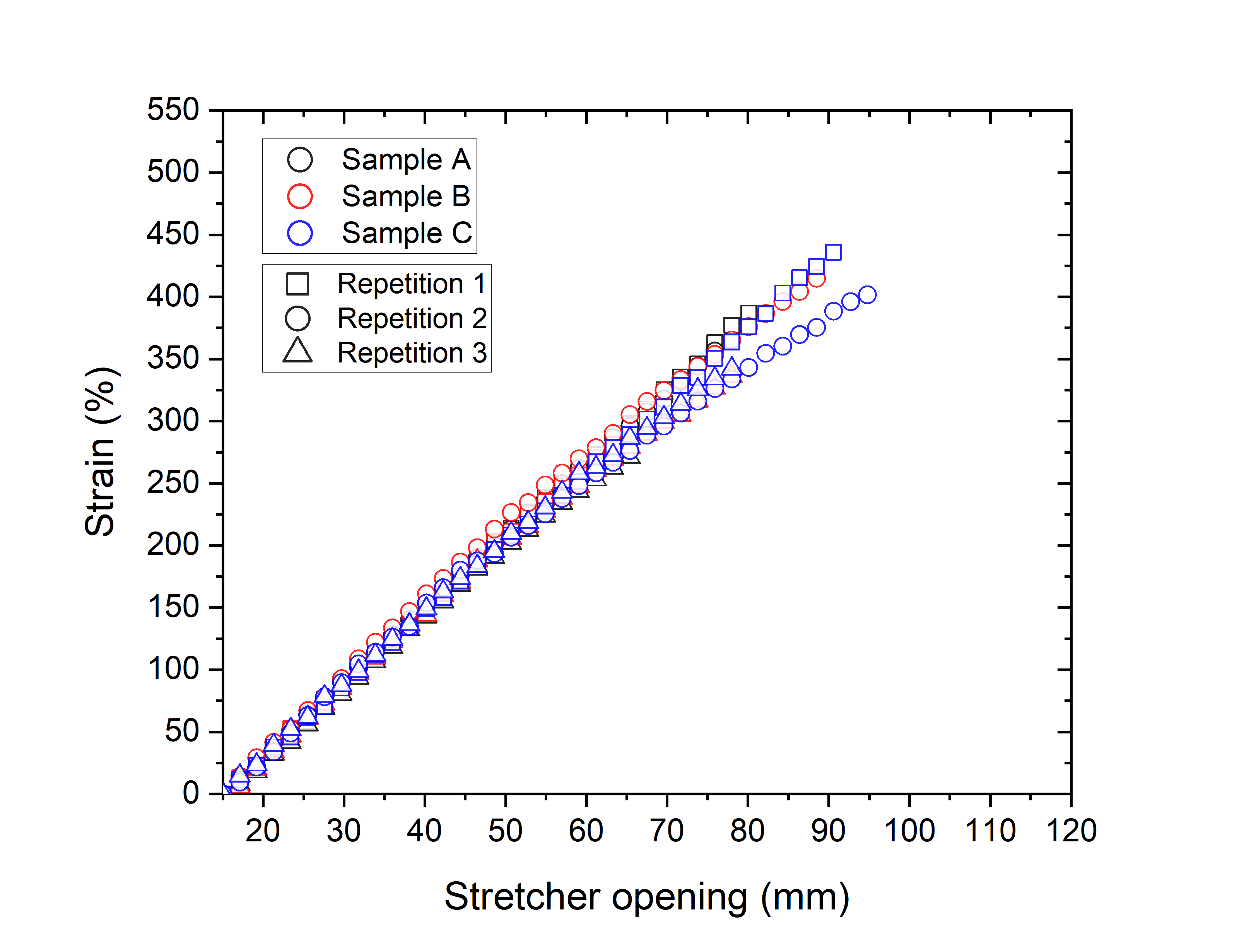}}
\caption{Relationship between stretcher opening and visually measured strain (distance between reflecting dots): the almost linear relationship between stretcher opening and strain confirms the reliability of the anti-slip patterns milled to the clamps that hold the samples.  
\label{strs-strn}}
\end{figure*}

Rubber and its mixtures are a vital component in industrial use. The versatility of this material allows it to be utilized in almost all known methods of transportation (such as cars, planes, boats, trains, etc.), in the construction of buildings, medical and sports devices, and general consumer products. Within such a wide range of applications, its main advantage is the viscoelastic behavior which allows the creation of products with a high degree of flexibility and high resistance to deformations.

Nevertheless, even the most resistant rubber composites are not perfect: such factors as aging, temperature variation,\textsuperscript{\cite{demassieux_temperature_2019}} wear and fatigue under stress,\textsuperscript{\cite{heinrich_cavitation_2020, lefevre_cavitation_2015, chang_cavitation_2001}} or inadequate reinforcement/mixing inevitably introduce damage, that starts from the cavitation on molecular and sub-molecular levels. The presence of cavitation phenomena poses a significant challenge across the entire industry. It often shortens the life span of the final product and negatively affects desired characteristics. Tires are especially affected by cavitation due to the technologically complex production process, harsh usage environment, and extremely high loads during the performance on the road. Therefore, there has been an ongoing industrial pursuit for the method of detection, evaluation, and quantification of cavitation, to achieve the final product with optimized qualities.

A multitude of methods are usually used for the evaluation of cavitation. One of the oldest and most popular methods for evaluating cavitation, which also provided the theoretical basis for the phenomena, is tensile testing.\textsuperscript{\cite{chang_cavitation_2001, heinrich_cavitation_2020, li_constitutive_2007, gent_cavitation_1990, gent_compression_1959}} Although this method allows the collection of statistically reliable descriptions of the physical properties of the rubber compounds very fast, direct observation of what is happening in the bulk of the material on a nanometer scale is beyond its scope. Dilatometry was reported to be used to validate established models.\textsuperscript{\cite{le_cam_review_2010}} Microscopy methods, including TLM, TEM, SEM, AFM,\textsuperscript{\cite{gigante_analysis_2021,bakhshizade_evaluation_2023, ao_fracture_2016, le_cam_importance_2006, gdoutos_nucleation_2006}} and dynamic image correlation are often used together with tensile tests.\textsuperscript{\cite{le_cam_volume_2008, federico_resolving_2021}} While dynamic image correlation provides insight into the multi-axial behavior of the sample, microscopy methods offer a micron-range resolution to observe the cavitation. However, sample preparation is necessary and these methods enable only the observation of the surface of the rubber rather than its internal structure. Micro-computed X-ray tomography ($\mu$CT) is often used to access the bulk of rubber composites.\textsuperscript{\cite{euchler_first-time_2021, euchler_situ_2020, heinrich_cavitation_2020, federico_resolving_2021, gdoutos_nucleation_2006}} It allows for analyzing the internal arrangement of rubber composites on a volumetric scale and observing the structural origin of volume changes in 3D. However, it requires reasonably fast scanning of the sample and can only deliver this information within the limits of the available spatial resolution.

Small-angle X-ray Scattering (SAXS), on the other hand, is a proven method that can describe hierarchical structures over several orders of magnitude within a single scattering image. Therefore, it is often used to characterize the bulk hierarchical structure of the filler within the rubber and follow structural changes in the sample during \textit{in situ} experiments on multiple length scales at the same time.\textsuperscript{\cite{staropoli_hierarchical_2019, robbes_situ_2022, okoli_dispersion_2022, yamaguchi_hierarchically_2017, baeza_multiscale_2013, shui_how_2021, takenaka_analysis_2013, koga_new_2008, mcglasson_quantification_2020, bouty_interplay_2016, hashimoto_hierarchically_2019, rishi_dispersion_2020}} The technique has few requirements for sample preparation and almost all shapes standardized by ISO for rubber evaluation can be used for the measurements. 
SAXS has already been explored as a means to characterize cavitation within rubber compounds.\textsuperscript{\cite{demassieux_temperature_2019, beutier_situ_2022, xiang_competition_2022, cao_nanocavitation_2018, zhang_nanocavitation_2012, zhang_opening_2013}} However, the previous research was limited by instrumental restrictions on available length scales, and generally low throughput of the setup, used for measurements.

In this work, we use for the first time the unique capabilities of Ultra Small-Angle X-Ray Scattering (USAXS) as provided by the beamline ID02 at the European Synchrotron Radiation Facility (ESRF) in Grenoble, France to study during \textit{in situ} stretching both cavitation effects and structural changes in rubber on the length scale from nanometers to tens of microns, simultaneously.\textsuperscript{\cite{narayanan_performance_2022}} For high throughput, a new sample stretcher has been developed. By utilizing the state-of-the-art set-up, we were able to obtain new information related to cavitation phenomena and discuss a comprehensive approach to the analysis of the recorded data.

\section{Materials and Methods}\label{MM}

A custom-made device has been designed for \textit{in situ} stretching of rubber, which enables high sample throughput. Ultra Small-Angle X-ray Scattering is used to follow the cavitation phenomena and structural changes in the bulk of the rubber samples during the application of strain. Samples with systematically varying compositions have been produced to validate the performance of the proposed experimental set-up and analysis.

\subsection{Custom-made stretching device}\label{strain-mes}

To ensure robust material characterization with a statistical approach that meets industrial standards, a special (U)SAXS stretching device was developed to observe structural changes, occurring cavitation, and the breaking of samples during the application of strain (Figure~\ref{setup}). The in-house developed sample stretcher enables accurate clamping between two metal plates with engraved anti-slip patterns that increase the effective surface area and reduce sample slipping during measurements. Our approach involves the repeated measurement of multiple individual samples of the same material to ensure statistical robustness.

The device offers the capability to simultaneously clamp and measure up to 15 samples with a maximum opening of 120 mm. For the experiment described in this article, the initial sample length was fixed at 15 mm in order to apply strain theoretically up to 700\%. The strain was applied in 50 steps (of 2.1 mm each) from 15 mm to the maximum opening of 120 mm. At each strain level, the experimental stage was moved horizontally and SAXS patterns were registered at the central part of each installed sample. Measurement of one strain state of the fully loaded stretching device required approximately 1 minute 40 seconds. Therefore, we consider our measurements quasi-static. All 50 strain states were measured in about 1.5 hours.

In order to observe the possibility of the sample slipping in the clamping mechanism during the experiments and to correct it if necessary, we have introduced a method using light-reflecting markers. These markers were applied to the center of the sample surface at an initial distance of 10 mm from each other and tracked optically by a small monochrome camera. In this way, we could track the actual deformation (strain) of the samples during the stretching experiments and guarantee precise and reliable strain measurements throughout the stretching process. Figure~\ref{strs-strn} shows the evolution of the measured strain by analyzing the movement of the reflecting dots as a function of the nominal opening of the stretching device for 3 different samples and 3 measured repetitions for each sample. The good linearity of the curves proves that sample slipping can be almost neglected. In any case, for maximum accuracy, all strain values referred to in this article are based on the optically measured evolution of the reflecting markers.

\begin{figure*}[!b]
\centerline{\includegraphics[width=0.85\textwidth]{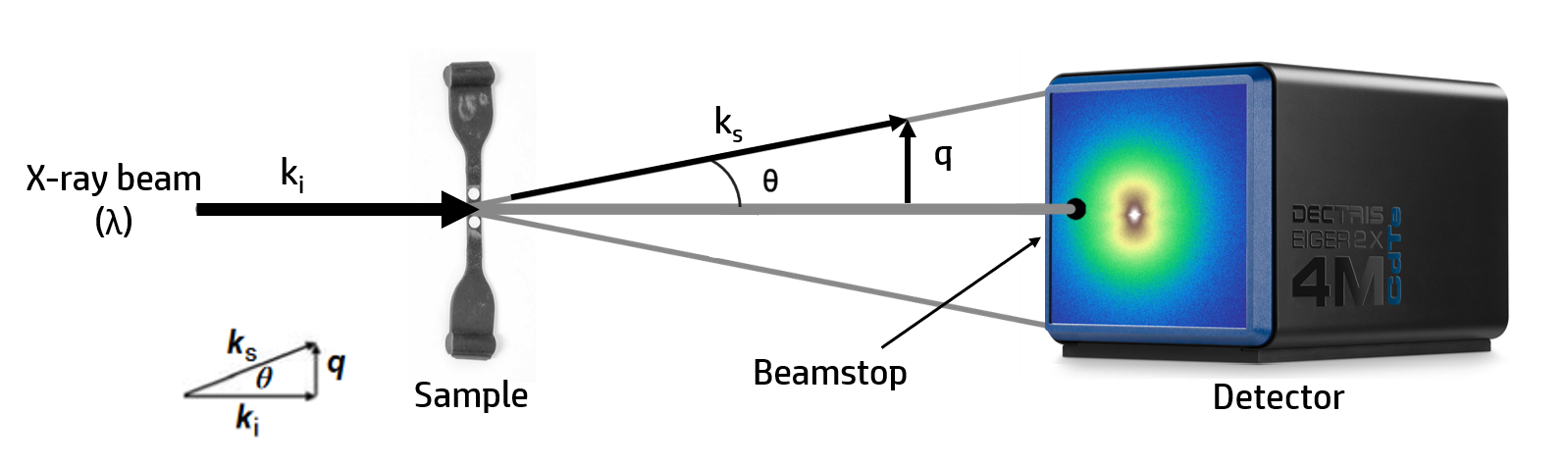}}
\caption{Schematic layout of a (U)SAXS set-up depicting the incident ($ \protect\boldsymbol{k}_i $) and scattered ($ \protect\boldsymbol{k}_s $) wave vectors, transmitted beam, scattering angle $\theta$, the 2D detector, and the scattering vector ($\protect\textbf{\emph{q}}$) with $\protect\textbf{\emph{q}} =  \protect\boldsymbol{k}_s - \protect\boldsymbol{k}_i$, where $|\protect\boldsymbol{k}_s| = |\protect\boldsymbol{k}_i| = 2\pi/\lambda$} \label{fig1}
\end{figure*}

\subsection{(Ultra) Small-Angle X-ray Scattering measurements}\label{ID02}

The USAXS/SAXS measurements were carried out at the European Synchrotron Radiation Facility (ESRF) located in Grenoble, France. The Time-Resolved Ultra Small-Angle X-Ray Scattering (TRUSAXS) beamline ID02 was utilized to obtain 2D scattering patterns of the specimens using a state-of-the-art 2D photon-counting pixel array detector Eiger2 X 4M from Dectris, Switzerland.\textsuperscript{\cite{narayanan_performance_2022}} The X-ray wavelength for the measurements was $\lambda$ = 0.101 nm. The set-up was collimated in the pinhole configuration. The incoming beam was attenuated using a 50 $\mu$m Molybdenum filter due to the high scattering of samples. The experiments were performed at room temperature with a variable sample-to-detector distance of 1 m, 8 m, and 31 m to cover the entire available range of the scattering vector \textit{q}: 
\begin{align}\label{eq1}
    \textit{q} &= |\textbf{\emph{q}}| = \frac{4\pi}{\lambda} \cdot \sin\left(\frac{\theta}{2}\right) 
\end{align}

from 0.0016 nm\textsuperscript{-1} to 7 nm\textsuperscript{-1}. The set-up used a circular beam stop of 1 mm in diameter for 31 m distance and a circular beam stop of 3 mm in diameter for 8 m and 1 m (Figure~\ref{fig1}). The sample transmission was accurately recorded using a diode situated behind the entrance window of the evacuated detector tube.

The measured 2D scattering patterns were normalized to an absolute intensity scale, cleared from detector gaps, azimuthally regrouped, and averaged to one-dimensional scattering curves I(\textit{q}) or $I(\psi)$, with $\psi$ being the azimuthal angle, using the ID02 standard data reduction process and the $SAXSutilities~2$ software package.\textsuperscript{\cite{boesecke_saxs_nodate, michael_saxsutilities2_2021}} A partial averaging was often applied as discussed in more detail in section \ref{res}. All measured (U)SAXS data was normalized to mm\textsuperscript{-1}, taking into account the reduction in sample thickness due to the stretching of the sample. The correction factor for the sample thickness is proportional to the sample transmission according to the Lambert-Beer law:

\begin{align}\label{eq2}
I_1=I_0~exp(-\mu t)\
\end{align}

with $I_0$ and $I_1$ being the incident and transmitted intensity, $\mu$ being the absorption coefficient of the sample, and t the sample thickness.\textsuperscript{\cite{alsnielsen_elements_2011}} Assuming a constant $\mu$ within the sample, the transmission coefficient $T$:

\begin{align}\label{eq3}
    T = \frac{T_1}{T_0}
\end{align}

has a power law dependence on the change of the sample thickness during the stretching, and therefore allows calculating the sample thickness from the measured change in the transmission:

\begin{align}\label{eq4}
    t_{n\%} = \frac{\ln(T_{n\%})}{\ln(T_{0\%})} \cdot t_{0\%}
\end{align}

with $n$ indicating the strain applied to the sample.

\subsection{Samples}\label{samples-desc}

\begin{table*}[!t]
\caption{Recipes of samples (in Vol\%).\label{Recipes}}
\newcolumntype{C}{>{\centering\arraybackslash}X}
\begin{tabularx}{\textwidth}{CCCC}
\toprule
\textbf{Components}		& \textbf{Sample A}& \textbf{Sample B} & \textbf{Sample C}\\
\midrule
SSBR		& 37.68			& 44.42		& 54.50	\\
Silica	&26.21	&25.05	&22.80\\
TDAE	&27.96	&21.97	&13.48\\
6PPD	&0.84	&0.99	&1.22\\
Wax	&0.96 &1.14	&1.39\\
ZnO	&0.19	&0.22	&0.27\\
Steric acid	&0.83	&0.98	&1.21\\
TESPD	&3.83	&3.61	&3.32\\
DPG	&0.60	&0.56	&0.52\\
TBBS	&0.55	&0.65	&0.79\\
Soluble sulfur 	&0.35	&0.41	&0.50\\

\bottomrule
\end{tabularx}
\end{table*}

\begin{table*}[!t]
\caption{Results of additional standard measurements performed at Continental AG.\label{addons}}
\newcolumntype{C}{>{\centering\arraybackslash}X}
\begin{tabularx}{\textwidth}{CCCC}
\toprule
\textbf{Measurement type}& \textbf{Sample A}& \textbf{Sample B} & \textbf{Sample C}\\
\midrule
Stress at 300 \% strain (M300) & 6.2 MPa& 6.9 MPa& 12.0 MPa	\\
Hardness&69.5 Shore A&71 Shore A&71 Shore A\\

\bottomrule
\end{tabularx}
\end{table*}

Three types of rubber compounds were provided by the European automotive parts manufacturing company Continental AG, located in Hanover, Germany. The basic idea behind the sample composition was to have controlled variations of the three main components of the rubber: polymer matrix (styrene butadiene rubber (SBR)), silica (HD silica SA 160), and oil (treated distillate aromatic extract (TDAE)). The laboratory compounds were mixed using an internal mixer applying a standard mixing protocol. In the first mixing step, rubber, filler, silane, and chemicals are added – the mixing time is sufficiently long to allow a chemical reaction between silane and silica. Accelerator and sulfur are added in the final mixing step. Rubber samples have been cured at 160°C for 20 minutes. Recipes of samples and their respective names can be found in Table \ref{Recipes}. Values for additional standard measurements, performed at the laboratories of Continental AG for the sample systems can be found in Table \ref{addons}. The hardness was measured according to DIN 53 505 and the tensile test was performed on a universal testing machine according to DIN 53 504. The former represents a standard measure for the static stiffness of a rubber compound and is within 1.5 Shore A points very similar between all three samples. From the tensile test, the stress at 300 \% strain (often referred to as M300) is reported in Table \ref{addons} and it is apparent that with increasing amounts of oil and filler, the M300 is significantly decreasing and hence the overall material stiffness is reduced.

Sample A has a significantly reduced amount of the polymer in relation to the filler components: 37.6 vol \% of polymer versus 26.2 and 27.9 vol \% of silica and oil; the relation of the polymer to silica and oil is 0.695. Sample B has a larger SSBR volume with 44.4 vol \% of polymer versus 25.0 and 21.9 vol \% of silica and oil; the relation of the polymer to silica and oil is 0.945. Sample C represents the system where the polymer amount dominates over the filler amounts within the composite: it has 54.5 vol \% of polymer versus 22.7 and 13.4 vol \% of silica and oil; the relation of the polymer to silica and oil is 1.5.

All dumbbell-shaped specimens were provided for the X-ray study in a state without any fatigue history and measured using the stretching device described in section \ref{strain-mes}. The geometrical parameters of the samples correspond to American Society for Testing and Materials (ASTM) standard D4482-11(2021).\textsuperscript{\cite{d11_committee_test_nodate}}

\section{Quantitative analysis}\label{res}

\begin{figure*}[!ht]
\centerline{\includegraphics[width=11 cm]{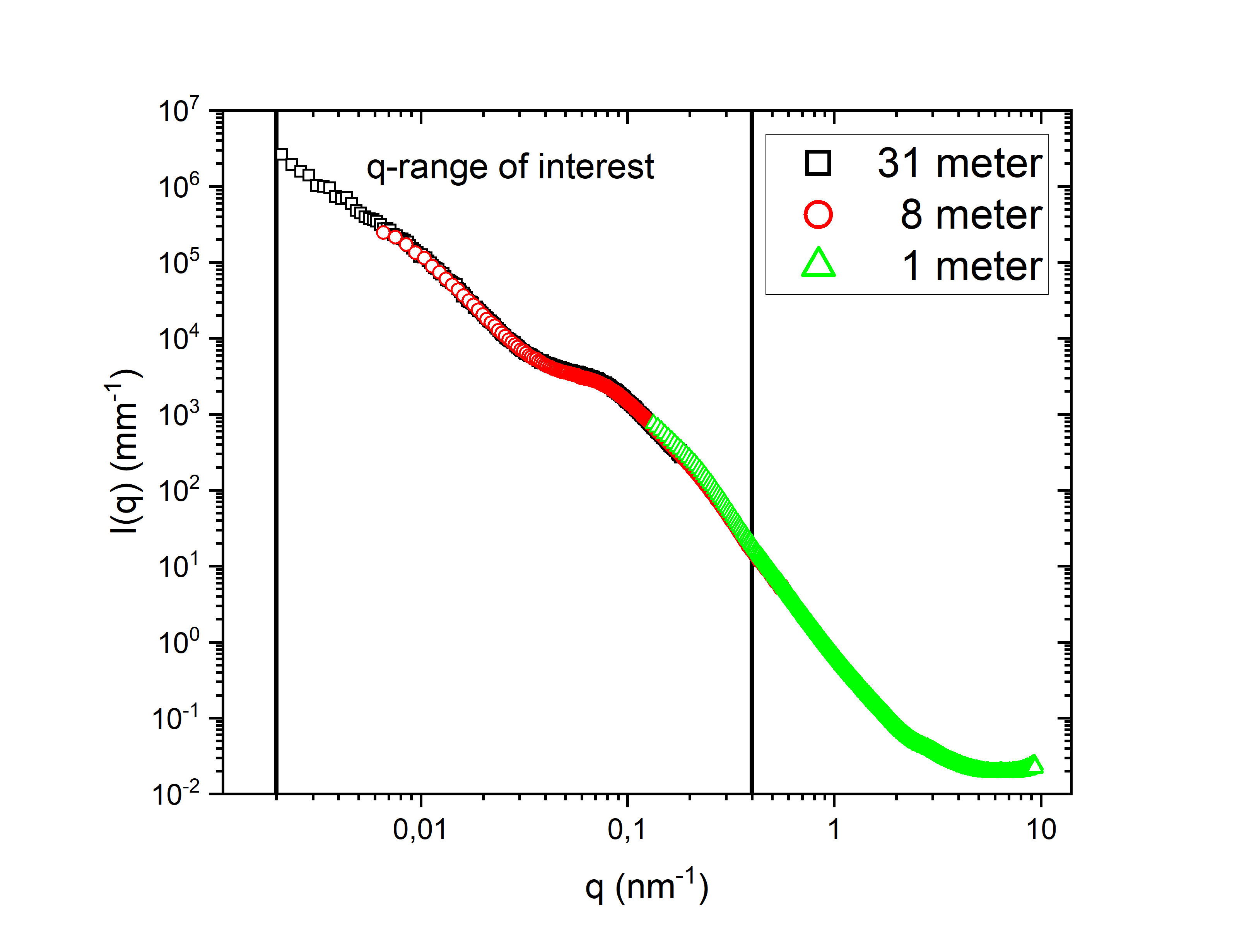}}
\caption{I(\textit{q}) of \textbf{Sample B} at 0\% strain: comparison of the scattering intensity partially averaged in the vertical direction obtained at 3 different sample-to-detector distances of 31 m, 8 m, and 1 m at the ID02 beamline. The \textit{q}-range, where most of the structural changes (including cavitation) are happening is limited by vertical black lines and mostly covered by measurements at a 31 m sample-to-detector distance. \label{merge}}
\end{figure*}

USAXS/SAXS patterns have been recorded over the available \textit{q}-range between 0.0016 nm\textsuperscript{-1} and 7 nm\textsuperscript{-1} (Figure~\ref{merge}) at 3 sample-to-detector distances (31 m, 8 m, 1 m). The scattering at high \textit{q} (above 1 nm\textsuperscript{-1}) is dominated by scattering from the polymer matrix and is not further analyzed in this article. The signal originating from primary silica particles is visible in the \textit{q}-range from 0.1 nm\textsuperscript{-1} to 1 nm\textsuperscript{-1}. In the following analysis, we focus on the analysis of the USAXS regime between 0.0016 nm\textsuperscript{-1} and 0.4 nm\textsuperscript{-1} where we can observe the hierarchical structure of silica filler particles, their clusters, and fractal branched structures. Moreover, in a given range the onset of cavitation can be observed, and the evolution of the number density, and size of cavities can be determined. This \textit{q}-range is mostly covered by measurements taken at 31 m sample-to-detector distance with some complementary information from data recorded at 8 m (Figure~\ref{merge}, black lines).

\begin{figure*}[!ht]
\centerline{\includegraphics[width=0.95\textwidth]{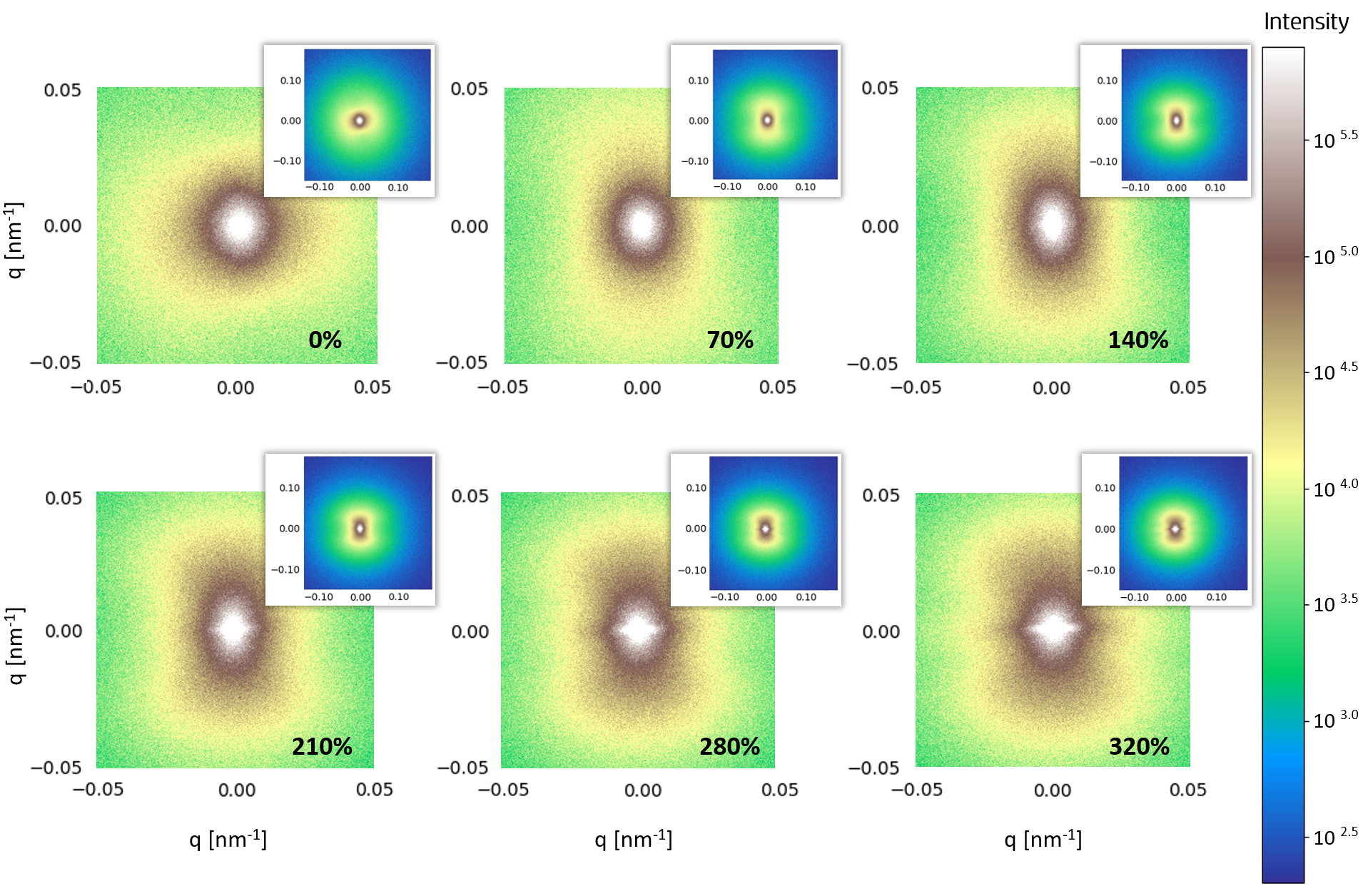}}
\caption{Magnification of the evolution of scattering patterns at ultra-small angles with the applied stain as indicated in the scattering images for the \textbf{Sample B}. The insets show the full scattering pattern recorded at a sample-to-detector distance of 31 m in a logarithmic intensity scale. The initial orientation of the scattering patterns (observed at 0\% strain, top left) gets oriented with increasing strain. The strain was applied in the vertical direction. An intensity streak in the horizontal direction starts to appear from 140\% strain, indicating the appearance of cavities. \label{patterns}}
\end{figure*}

Already in the absence of strain (Figure~\ref{patterns}, top left), the scattering of rubber compounds displays an oriented alignment often referred to as a "butterfly" pattern.\textsuperscript{\cite{staropoli_hierarchical_2019,robbes_situ_2022,beutier_situ_2022, cao_nanocavitation_2018, zhang_opening_2013, zhang_nanocavitation_2012}} The pre-orientation is explained by the rearrangement of silica clusters and interpenetrating mass-fractal agglomerates during standard mixing procedures and vulcanization.\textsuperscript{\cite{staropoli_hierarchical_2019}}

Upon stretching, the pre-oriented USAXS "butterfly" pattern reorients in the direction of the applied strain and the scattering intensifies as the strain enforces the alignment of the filler network within the rubber to follow the vector of applied tension (Figure~\ref{patterns}, top row 0\%, 70\%, and 140\%). The reorientation effect has been previously observed in various investigations.\textsuperscript{\cite{staropoli_hierarchical_2019,robbes_situ_2022,beutier_situ_2022, cao_nanocavitation_2018}}

When the applied strain goes beyond typically 140\% of the original length we start to see noticeable changes happening in the patterns. A streak, perpendicular to the strain direction appears and rapidly grows in intensity (Figure~\ref{patterns}, bottom row), superimposing over the "butterfly" scattering of the filler. Prior studies have reported similar effects, associated with refraction phenomena occurring at elongated oriented interfaces within the material's bulk.\textsuperscript{\cite{hosemann_refraction_1987, stribeck_x-ray_2007, chang_retardation_2021,chang_critical_2018,chang_influence_2018}} Based on these previous findings, it can be concluded that the origin of the observed streak is the interaction of X-rays with interfaces within the bulk of the material, like the walls of cavities opened by the applied strain - whereas the "butterfly" patterns originate from elastic scattering of X-rays from the silica filler network.

For the quantitative analysis of the observed phenomena, we propose two methods. The first one analyzes the azimuthal dependence of the scattering intensity under gradually rising strain. The second method focuses on analyzing the \textit{q} dependence of the scattering signal at different strain levels in the direction parallel and perpendicular to the applied strain. By employing both methods, we gain complementary information about cavitation as a function of strain and related structural changes within the rubber matrix. 

\subsection{Analysing the azimuthal dependence of the scattering intensity in the USAXS regime}\label{azimuthal}

By looking into the scattering patterns as a function of the azimuthal angle we aim to determine the earliest strain state where a streak originating from the cavitation can be observed. The main difficulty at this step is that the onset of streak formation has to be detected while being superimposed on an evolving scattering pattern originating from the alignment of the silica filler particles.

\begin{figure*}[!ht]
\centerline{\includegraphics[width=0.8\textwidth]{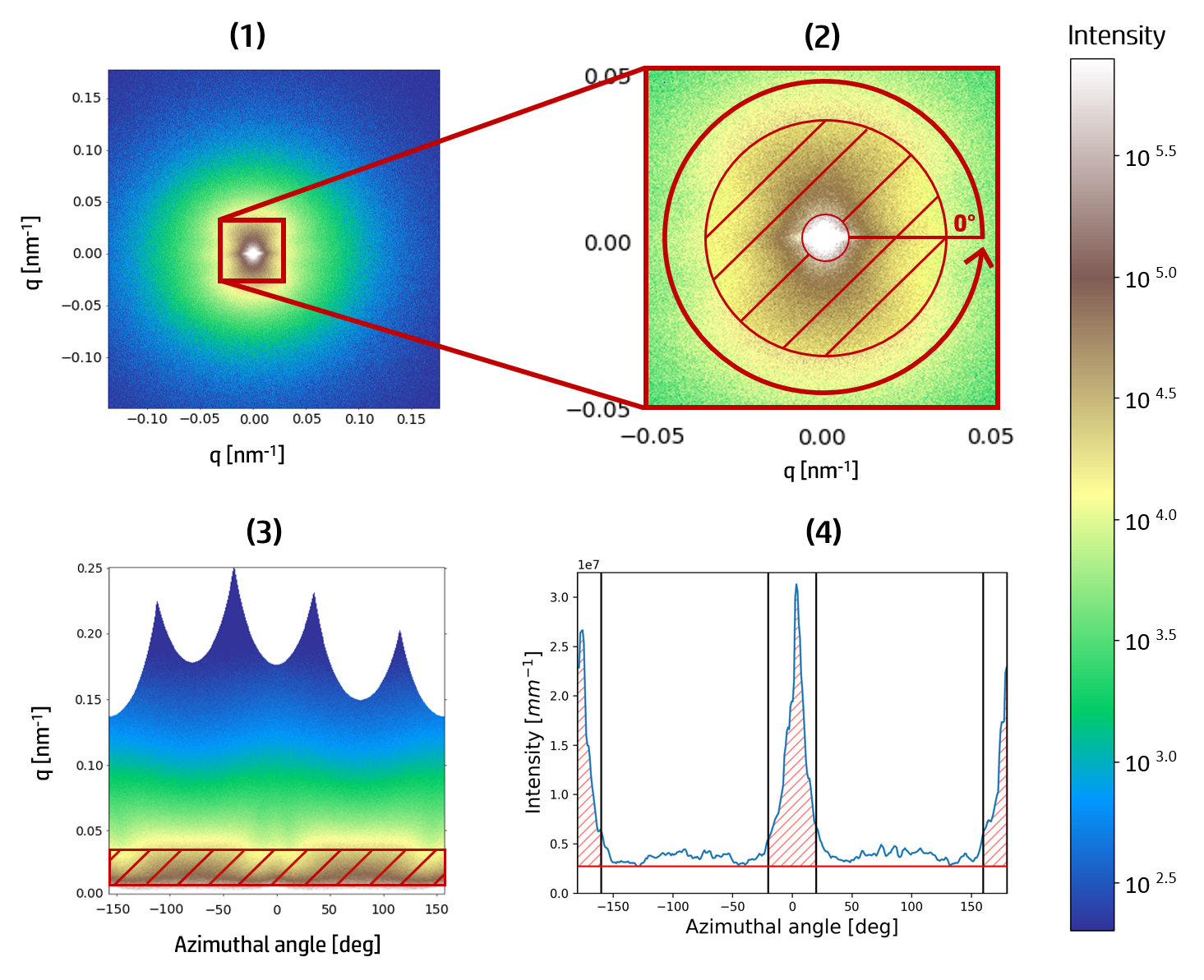}}
\caption{Procedure for the evaluation of the streak intensity for \textbf{Sample B} at 320\% strain (sample strained vertically): \textbf{(1)} - full scattering pattern recorded at a single sample-to-detector distance of 31 m; \textbf{(2)} - zoom in the central area: the arrow indicates the azimuthal angle, the hatched area indicates the \textit{q}-range between 0.0018 nm\textsuperscript{-1} and 0.0035 nm\textsuperscript{-1} which was used for averaging along the scattering vector \textbf{q}; \textbf{(3)} - azimuthally regrouped representation of the scattering image (1) with the \textit{q}-range used for averaging (hatched); \textbf{(4)} - averaged intensity as a function of the azimuthal angle. The streak indicating the presence of cavities is found at 0\textdegree and 180\textdegree in the horizontal direction (hatched). The red line indicates the level of scattering background, originating mostly from the filler particles, that is later subtracted from the obtained intensity profile. \label{data-red}}
\end{figure*}

\begin{figure*}[!ht]
\centerline{\includegraphics[width=\textwidth]{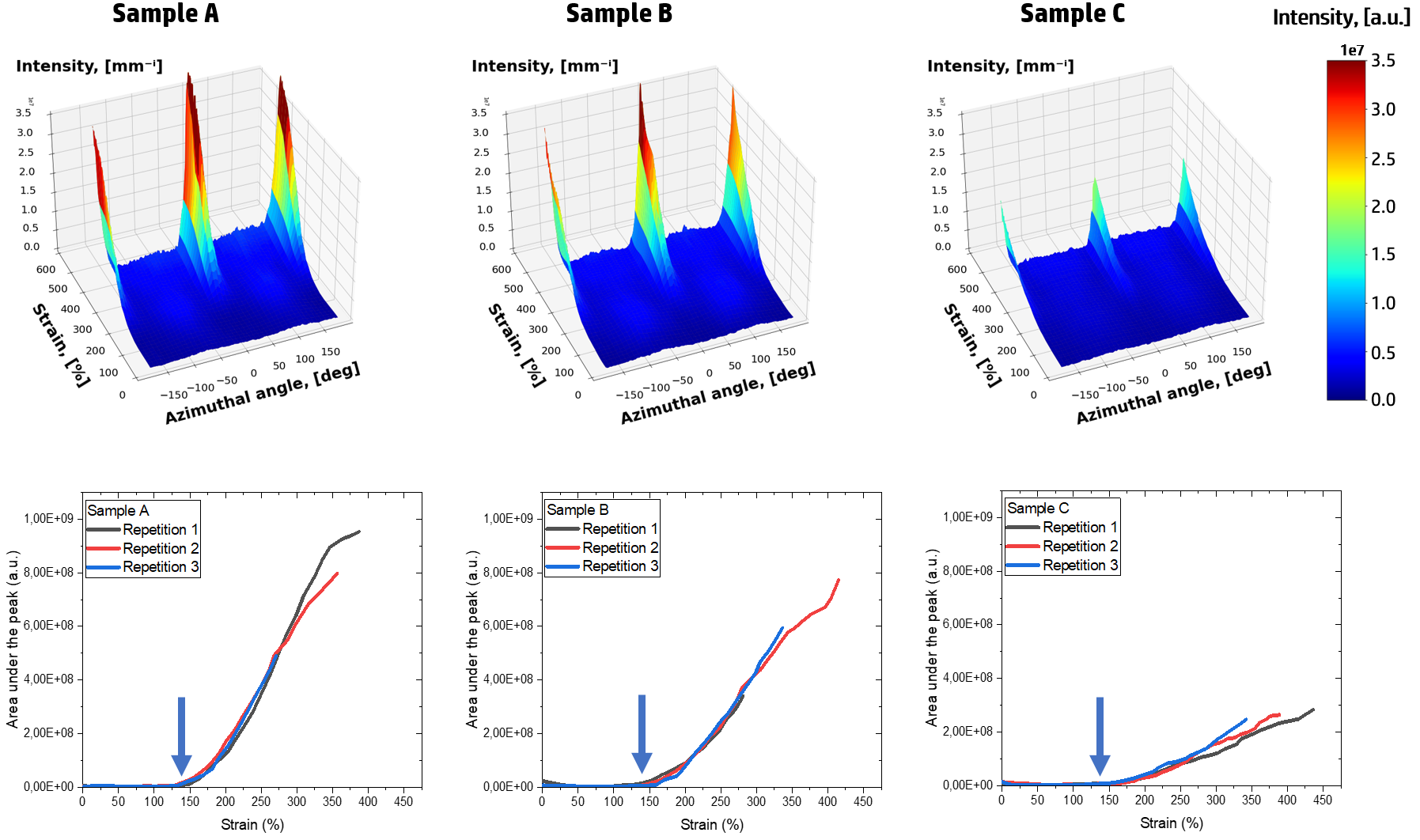}}
\caption{The top row shows 3D plots of the averaged intensity (determined by the procedure presented in Figure~\ref{data-red}) as a function of the azimuthal angle and strain for samples A, B, and C. The area under the streaks as a function of the strain for repeated sample measurements is presented in the bottom row. The rise of this integrated intensity during the strain application (indicated by the blue arrow) can be used as a precise measure for the onset of the cavitation in the samples. Different repetitions of stretching the same material give very reproducible results with an onset of 140\% for all samples reported in this article. On the other hand, the integrated intensity of the highest strain states decreases from Sample A to Sample C, indicating a lower number density of cavities created during the straining of the material. Due to sample breakage, not all strain values could be measured for some samples. \label{3d}}
\end{figure*}

As illustrated in Figure~\ref{data-red}, we average the measured intensity as a function of the azimuthal angle (indicated by the arrow in Figure~\ref{data-red} (2)) in a \textit{q}-range in the USAXS regime between 0.0018 and 0.0035 nm\textsuperscript{-1}. This averaging area is shown in Figure~\ref{data-red} (2) and (3) as a hatched area, where Figure~\ref{data-red} (3) is an azimuthally regrouped representation of (1). The resulting averaged azimuthal profile is shown in Figure~\ref{data-red} (4), identifying the streak at 0\textdegree and 180\textdegree. The observed streaks are oriented exactly perpendicular to the applied strain, due to the fact that most of the cavities are oriented in the direction of the pull. However, the streaks are superimposed on a strong scattering contribution originating from the filler particles. The two contributions are separated by searching for the minimum background value across the integrated scattering intensity and subtracting it from the resulting azimuthal profile (indicated by the red line in Figure~\ref{data-red} (4)). With increasing strain, we observe the streak progressively rising over the signal of the filler and we can determine precisely the moment when cavities appear. To better visualize the evolution of the streak formation, we stack the averaged azimuthal intensity profiles(Figure~\ref{data-red} (4)) as a function of the strain applied to our samples and create a 3D representation of our averaged intensity (Figure~\ref{3d} - top row).

By looking at the area under the streak (Figure~\ref{3d}, bottom row) as a function of the applied strain, we learn about the evolution of the number density of the cavities present in the probed rubber volume. The onset of cavitation is the point where the intensity of the streak area starts to rise. By looking at the resulting strain-area graphs, we discovered that the onset of cavitation in the rubber systems studied in this article stays the same at around 140\% for all three samples. The same onset value can be connected to the fact that all samples are comparable in the static stiffness usually measured at low deformation (i.e. hardness (Table~\ref{addons})). Hence, the formation of cavities at 140\% for samples A, B, and C points towards a balance between stiffness contribution from fillers on the one side and the polymer network on the other. This is not too unexpected because the samples have been designed to be balanced regarding these two contributors: an increasing filler amount is counterbalanced by an increased oil amount (decreasing the network contribution).

Beyond the indicated onset of cavitation, we also obtain the rate at which the number density of cavities evolves within the sample as the slope of the strain-area graphs. Here, the network contribution dominates the stiffness and therefore the potential resistance towards the evolution of damage. Thus the observed rate of cavity formation is significantly different between the three samples, which is clearly observable in Figure~\ref{3d} (bottom row). This is due to the fact that the internal stress distribution within the samples is significantly different due to the varying share of polymer within the volume of the sample. Since the polymer network has to carry most of the stress after the filler structures have been broken down, a sample with a reduced amount of polymer (e.g. sample A) will experience more stress per each supporting polymer bond. Its vulnerability to a higher number of damages is well shown in Figure~\ref{3d} by an increased growth slope.

\begin{figure*}[!ht]
\centerline{\includegraphics[width=0.8\textwidth]{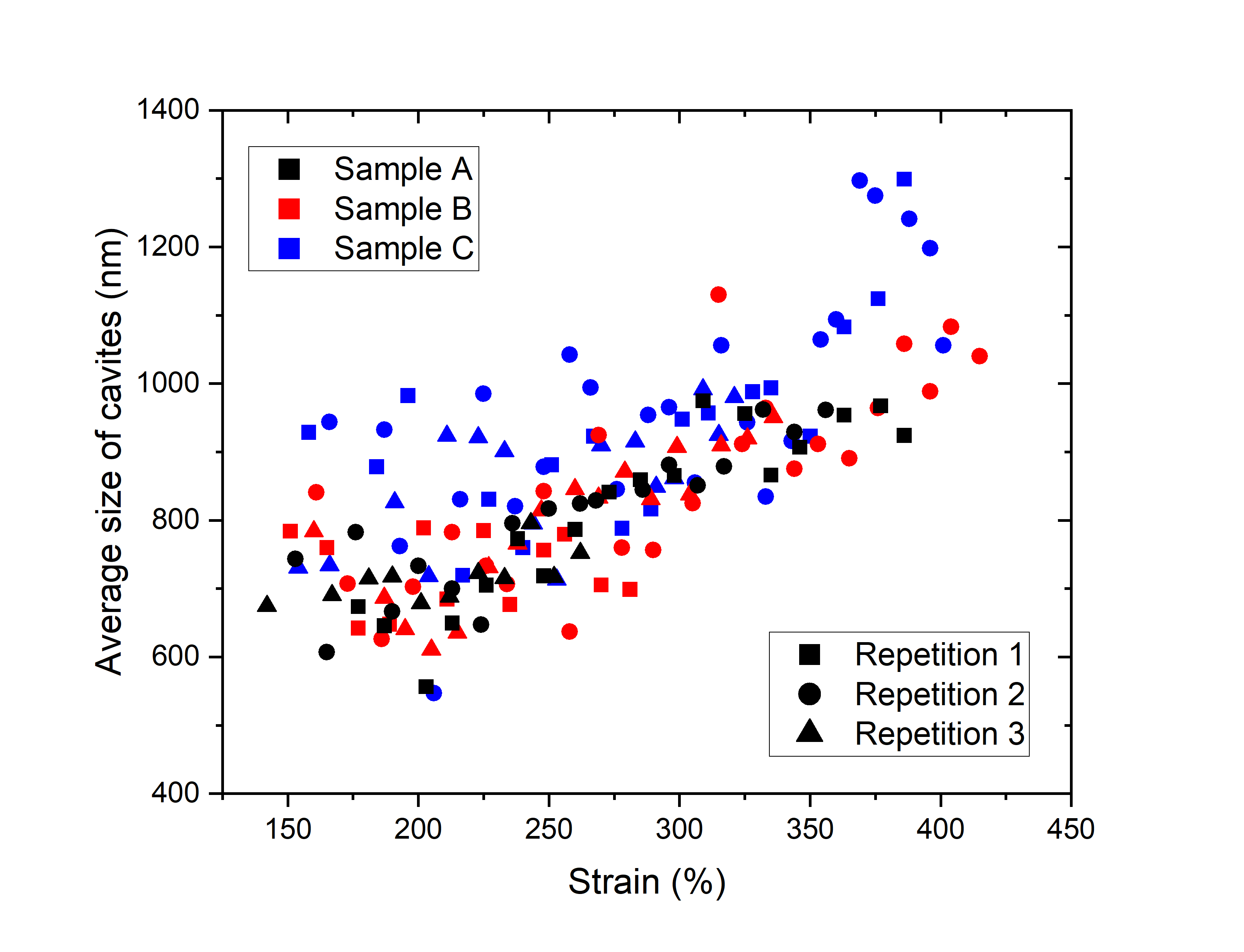}}
\caption{Calculated average size of cavities as a function of applied strain. Calculations were performed for multiple measurements of the same material and showed a very comparable trend. \label{cavities}}
\end{figure*}

To determine beyond the onset and cavitation rate also their size we apply the Ruland Streak Method. This method allows a description of the average longitudinal extension of cavities, rods, or micro-fibrils from streak-like scattering signals in case refraction is observed.\textsuperscript{\cite{stribeck_x-ray_2007,cao_nanocavitation_2018}} Refraction in our case comes from interfaces of cavities that are strongly oriented in the direction of the pull and creates an intensity streak that is well-oriented in the direction perpendicular to the strain. 
According to the method, the size $L$ of the cavities can be calculated from the measured integral breadth of the streak $B_{obs}$ as a function of the azimuthal angle:

\begin{align}
    L &= \frac{1}{q_\text{av} \cdot B_{\text{obs}}} \label{ruland}
\end{align}

where $q_\text{av}$ is the scattering vector where the angular cut was calculated - an average value in the USAXS regime between 0.0018 nm\textsuperscript{-1} and 0.0035 nm\textsuperscript{-1} for this study. The integral breadth was determined as the area of a peak divided by the peak height after subtraction of the scattering signal of the silica filler. Using this calculation we can obtain the average sizes $L$ of the cavities for each strain value where the streak is visible as presented in Figure~\ref{cavities}.

The evolution of the cavity size between Samples A, B, and C during stretching is very comparable and does not show differences as strong as in the previously discussed cavitation rate. Our samples exhibit a substantial difference in polymer concentration, however, the cavity size experiences only a slight increase in samples with a higher polymer fraction (Sample C). This observation leads us to the conclusion that the cavity size manifests as a more universally applicable effect of a polymer (SSBR), showing less dependence on the overall composition of the rubber compound. Consequently, the evolution of the number density of cavities visible through the cavitation rate emerges as a more crucial parameter for the comprehensive evaluation and comparison of samples.

By paying attention to the \textit{q}-dependent analysis presented in the following section, we will try to clarify the influence of cavity formation on the interconnected filler structures of the rubber compound.

\subsection{Analyzing I(\textit{q}) as a function of the scattering vector in the USAXS regime.}\label{scattering-classics}

\begin{figure*}[!t]
\centerline{\includegraphics[width=13 cm]{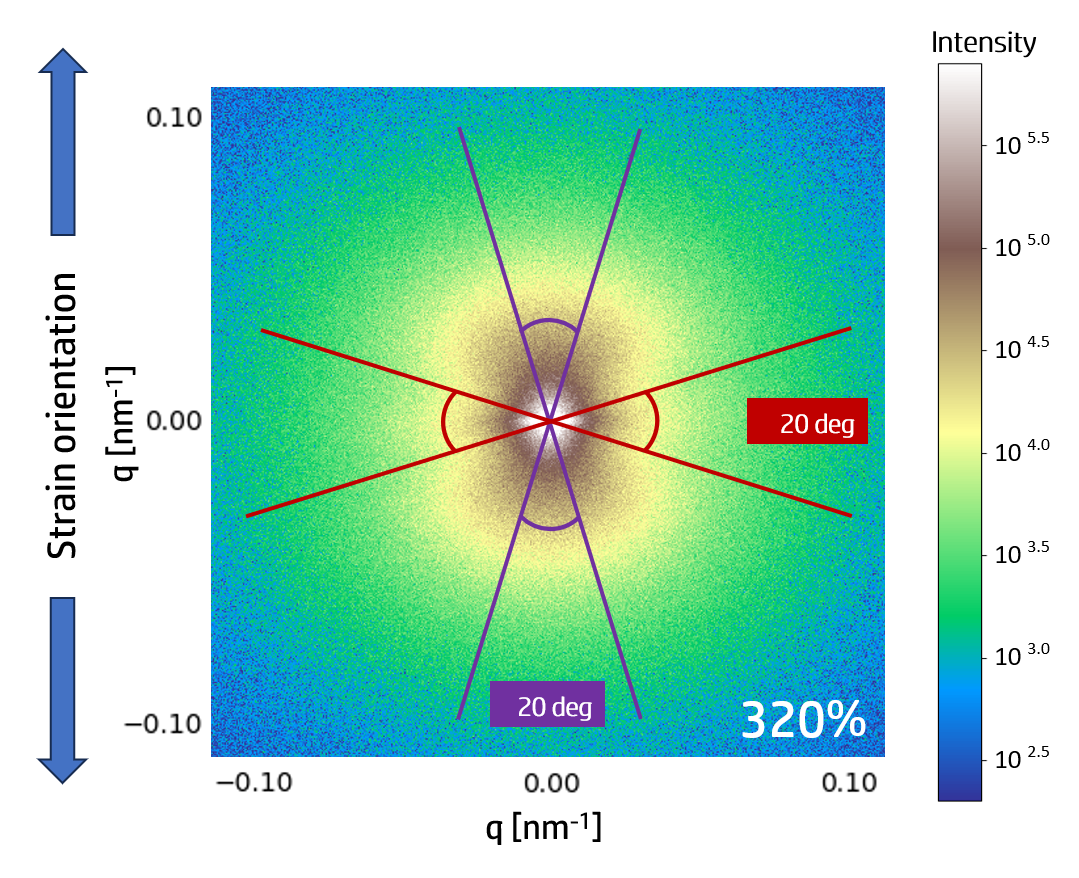}}
\caption{Partial integration scheme: 20 degrees integration ranges in directions parallel and perpendicular to the strain were calculated. In the vertical direction, only structural information from filled rubber is detected. In the horizontal direction refraction coming from the streak additionally contributes to the signal \label{oriented}}
\end{figure*}

In general, the scattered intensity as a function of the scattering vector I(\textit{q}) delivers valuable information about the hierarchical structure of the silica filler network within rubber compounds over many length scales. In the case of stretched rubber, two contributions are observed within the (U)SAXS scattering patterns. The first one is the butterfly-shaped scattering originating from the alignment of primary particles, clusters, and larger mass-fractal silica agglomerates. The second contribution is the refraction induced by cavities and observed perpendicular to the strain direction. It is superimposed on an evolving scattering of the filler particles as soon as the onset of cavitation is reached. Due to its preferential direction perpendicular to the applied strain, it can be (partly) separated from the scattering intensity by partially averaging the acquired SAXS patterns. 
Therefore, we focus here on two angular ranges for the azimuthal averaging of the scattered intensity as indicated in Figure~\ref{oriented}: The first one is centered $\pm$10$^\circ$ around the strain axis (purple range) and gives information of the strain-oriented hierarchical arrangement of the silica filler particles in that direction. It will be referred to as scattering in parallel direction ($\parallel$). The second angular range is centered $\pm$10$^\circ$ around the expected direction of the streak originating from the cavities (red range) and will be referred to as scattering in the perpendicular direction ($\perp$). Before the onset of cavitation, structural information about the arrangement of the silica filler network in that direction can be obtained. Beyond onset, the signal at higher strain levels is dominated by the refraction originating from the cavities.

To quantitatively describe the structures within the filled rubber compounds, we employ the Unified Scattering Function (USF) as a well-established method.\textsuperscript{\cite{beaucage_combined_2012}} The USF enables us to assess the dimensions, spatial distribution, and organization of the different-sized filler structures within the polymer matrix.\textsuperscript{\cite{staropoli_hierarchical_2019, robbes_situ_2022, okoli_dispersion_2022, yamaguchi_hierarchically_2017, baeza_multiscale_2013, shui_how_2021, takenaka_analysis_2013, koga_new_2008, mcglasson_quantification_2020, bouty_interplay_2016, hashimoto_hierarchically_2019, rishi_dispersion_2020,vilgis_reinforcement_nodate}} In rubbers, silica fillers are usually arranged hierarchically, creating a system composed of $n$ structural levels. In our analysis, we will follow the evolution of a system composed of 4 structural levels: the first level is located at high \textit{q} values and describes the primary silica particles. The second level, between 0.06 nm\textsuperscript{-1} and 0.15 nm\textsuperscript{-1}, corresponds to the clusters of silica primary particles. The third level describes structures of higher order observed in the \textit{q}-range between 0.004 nm\textsuperscript{-1} and 0.03 nm\textsuperscript{-1}. The fourth level was used to describe ramified filler agglomerates with tens of $\mu$m in size.

In USF, each described level is represented as a combination of Guinier and Power laws. For our compounds, we use the extended version of the function described by Beaucage (Equations \ref{eq:1} and \ref{eq:2}) with additional corrections by Hammouda (Equation \ref{eq:3}) that help us to follow an assumption of silica fractality within the rubber-embedded filler networks.\textsuperscript{\cite{vilgis_reinforcement_nodate,beaucage_approximations_1995,hammouda_analysis_2010}}

\begin{figure*}[!b]
\centerline{\includegraphics[width=\textwidth]{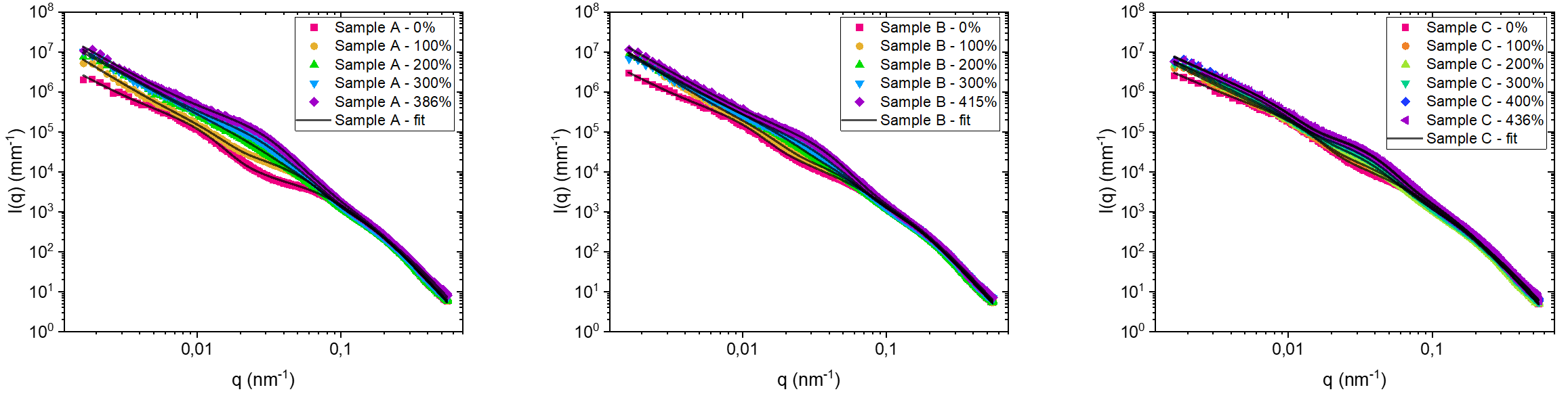}}
\caption{Evolution of the scattering curves for Samples A, B, and C in the direction parallel ($\parallel$) to the applied strain for different strain states as indicated in the legend. The continuous lines are unified fits using 4 structural levels. \label{fits-para}}
\end{figure*}

\begin{figure*}[!t]
\centerline{\includegraphics[width=\textwidth]{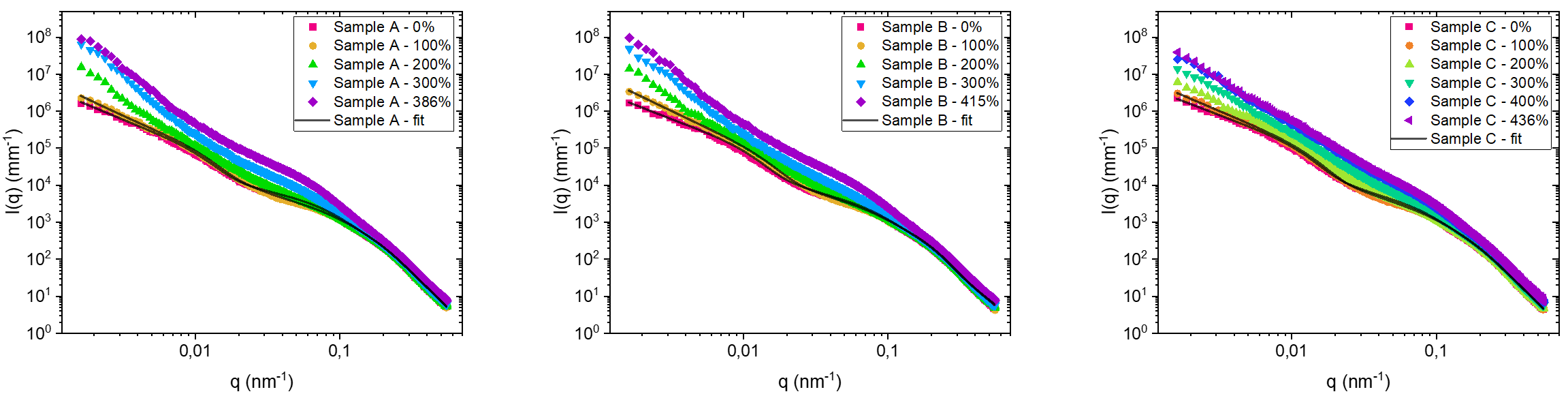}}
\caption{Evolution of the scattering curves for Samples A, B, and C in the direction perpendicular ($\perp$) to the applied strain for different strain states as indicated in the legend. The continuous lines are unified fits using 4 structural levels. \label{fits-perp}}
\end{figure*}

\begin{align}
    I(\textit{q}) &= \sum_{i=1}^{n} \left(G_{i} \cdot e^{-\frac{q^2 R_{g,i}^2}{3}} + B_{i} \cdot e^{-\frac{q^2 R_{g,i+1}^2}{3}} \cdot (q_i^*)^{-P_i}\right) \label{eq:1}
\end{align}

\begin{align}
    q_i^* &= \frac{q}{\left\{ \text{erf}\left(\frac{kqR_{g,i}}{\sqrt{6}}\right) \right\}^3} \label{eq:2} 
\end{align}

\begin{align}
    B_i &= G_i \cdot e^{-\frac{1}{2}P_i} \cdot \left(\frac{3P_i}{2}\right)^{\frac{P_i}{2}} \cdot \frac{1}{R_g^{P_i}} \label{eq:3}
\end{align}

In those equations, I(\textit{q}) is the scattering intensity, which represents the sum of contributions from $n$ structural levels within the rubber compound. These structural levels are characterized by their respective contrast values ($G$ and $B$), radius of gyration ($R_g$), and power law exponent ($P$). The radius of gyration ($R_g$) describes the characteristic size of the structural units, reflecting their spatial distribution within the material. 
\textit{q}\textsuperscript{*} in the equation involves the error function (erf), which enables a smooth transition between power law and Guinier regions. The parameter $k$ is empirical and has specific values depending on the nature of the structure, with approximately 1.06 for $P = 2$ and 1 for $P > 3$.

The size of the primary silica particles has been fitted from independent SAXS measurements of the raw silica material (not shown). Their radius of gyration $R_g$ was determined to be 11.5 nm. This size is used consistently in the USF fits of samples A, B, and C. For the fitting, we used scattering data recorded at 31 m and 8 m which covers well, after merging them in unique scattering curves, the relevant part of the full available \textit{q}-range (Figure~\ref{merge}, \textit{q}-range of interest). Fit parameters were determined independently for the partially averaged data in parallel $\parallel$ and perpendicular $\perp$ direction using the IRENA package in IgorPro.\textsuperscript{\cite{ilavsky_irena_2009}} Figures~\ref{fits-para} and \ref{fits-perp} show merged I(\textit{q}) curves in logarithmic scale with resulting fits in the corresponding partial integration ranges. It should be noted again that in the $\perp$ direction, the USF allows us to characterize scattering features of hierarchical structures in the scattering signal only until the refraction streak appears. Afterward, the contribution of the streaks dominates and does not allow any more to extract structural information.

Having a closer look at the evolution of the scattering curves with increasing strain in $\parallel$ direction (Figure~\ref{fits-para}) one observes that the scattering curves of all three samples do not show any evolution at high \textit{q} values. This is due to the fact that the scattering curves have been normalized for the decreasing sample thickness as explained before. Moreover, the size of the primary particles that contribute in this range is not expected to change.

More significant is the evolution of the scattering curves in the center of the plots. In the corresponding \textit{q}-range, we observe the signal of the silica clusters (second structural level in the USF). A clear shift of the Guinier level to the left is observed with increasing strain. This is explained by the fact that the silica clusters get elongated in the same way as the rubber is stretched in this direction. Further to the left another bump is observed in the scattering curves. This bump corresponds to the third structural level in the USF and is later discussed in the text. The rise of the scattering curves at the lowest \textit{q} values is due to large silica agglomerates which are, with a size of more than 10 $\mu$m, beyond the detectable size range of the instrument.

Looking at the scattering in the $\perp$ direction in Figure~\ref{fits-perp} one observes that again no change is observed at the level of primary particles. However, the Guinier level of the silica clusters moves only very little. A third structural level is detected further to the left as is the case in the $\parallel$ direction. The strong intensity increase from lowest \textit{q} values up to 0.1 nm\textsuperscript{-1} at strains of more than 140\% is explained by the appearance of the refraction streak. Therefore, data is not fitted anymore.

\begin{figure*}[!t]
\centerline{\includegraphics[width=0.9\textwidth]{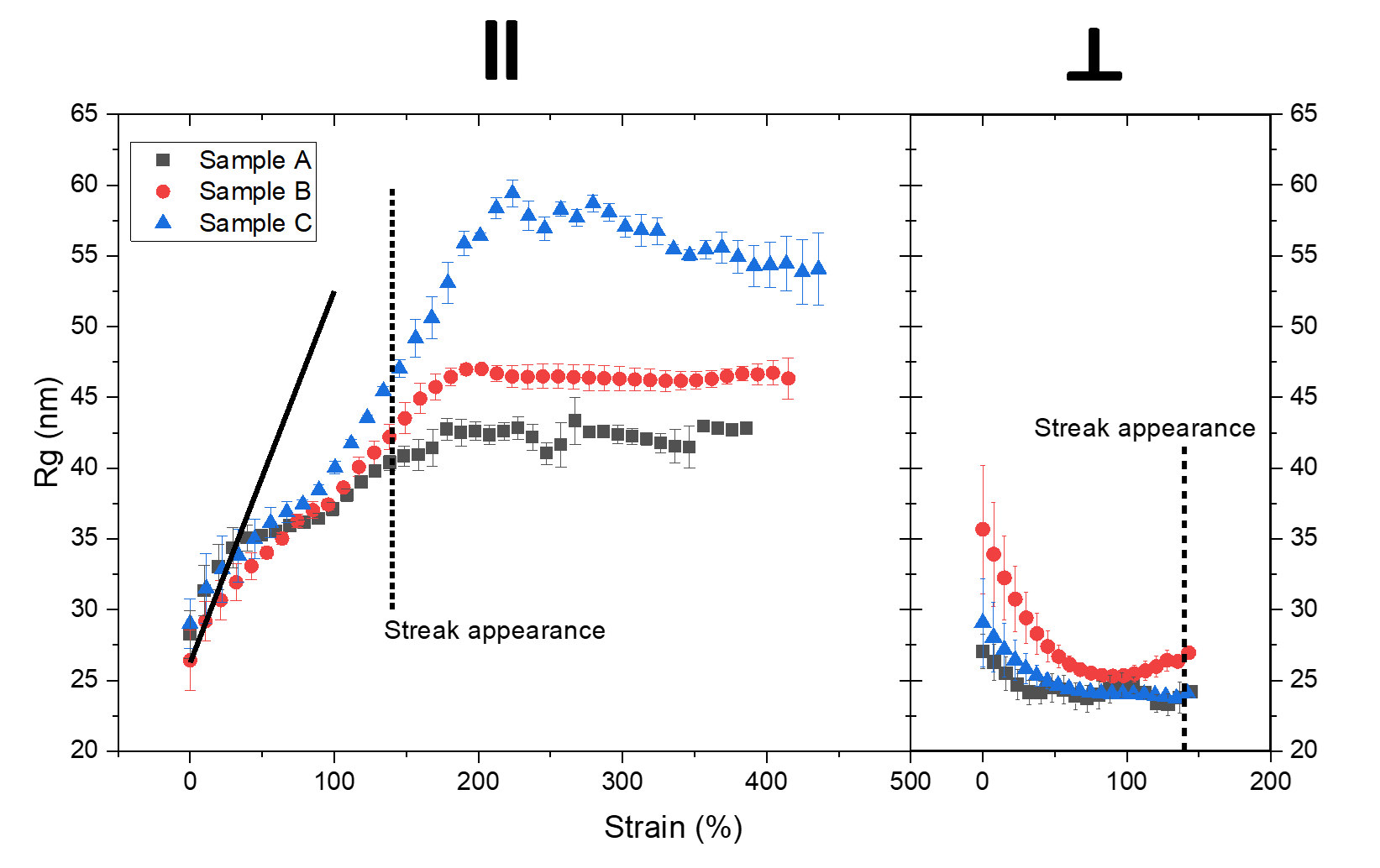}}
\caption{Radius of gyration of the silica clusters (second structural level) of Samples A, B, and C in the parallel and perpendicular direction to the applied strain. The solid black line indicates the theoretical elongation of the silica clusters assuming a pure affine deformation without any relaxation effects. The hashed line shows the moment when the streak is detected in the scattering patterns. Its presence does not allow the determination of a radius of gyration in the perpendicular direction beyond this point. \label{sizes-Rg2}}
\end{figure*}

\begin{figure*}[!t]
\centerline{\includegraphics[width=0.9\textwidth]{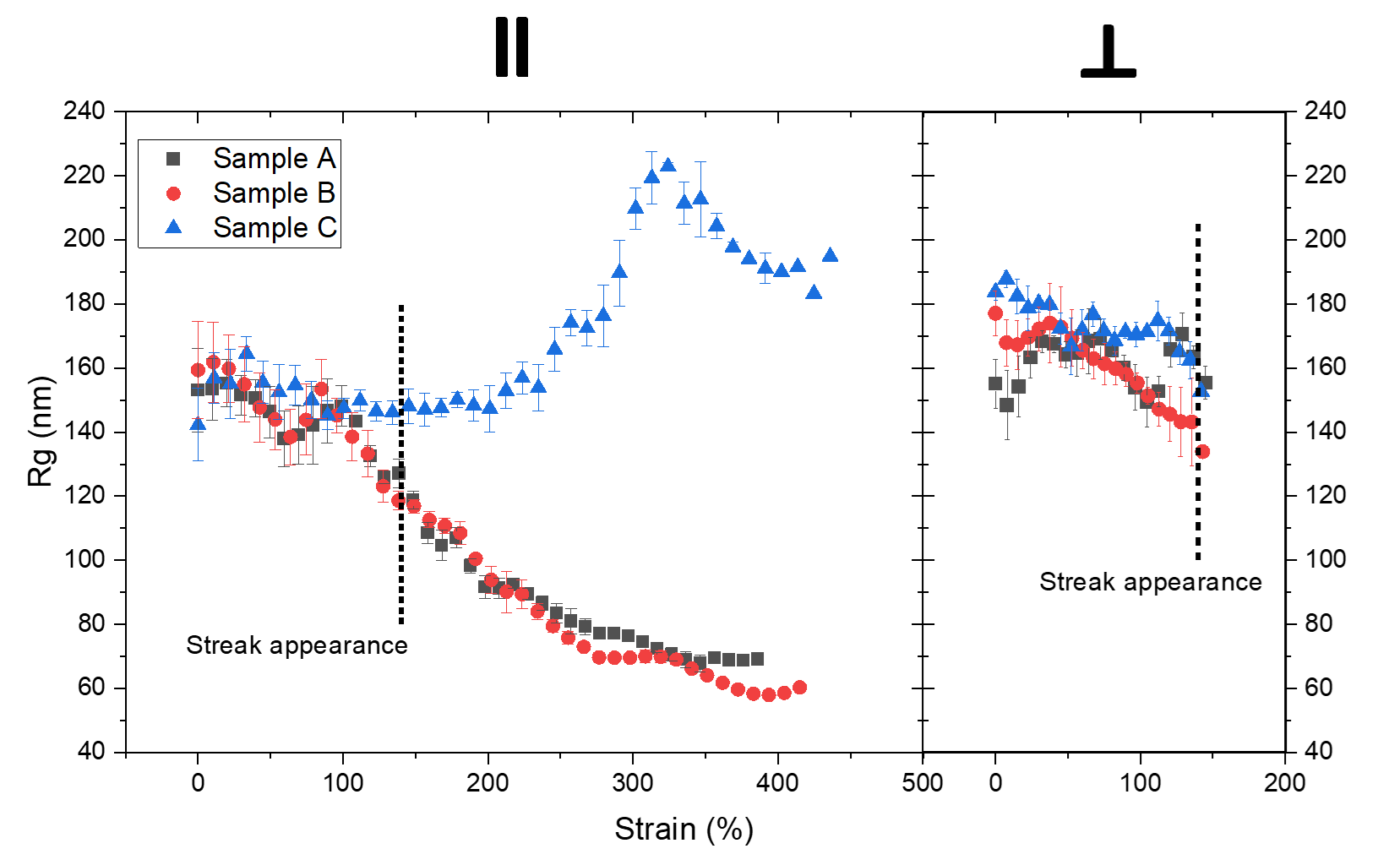}}
\caption{Radius of gyration of the structures observed in the third structural level of Samples A, B, and C in parallel and perpendicular direction to the applied strain. The hashed line shows the moment when the streak is detected in the scattering patterns. Its presence does not allow the determination of a radius of gyration in the perpendicular direction beyond this point. \label{sizes-Rg3}}
\end{figure*}

The detailed fit results of the radius of Gyration of the clusters are reported in Figure \ref{sizes-Rg2} for samples A, B, and C. 
During deformation, stresses from the polymer matrix are transmitted via the silica-polymer bonds to the silica clusters. Thus, there will be a balance between these intra-cluster forces and the forces at the cluster-polymer interface.
At low strain, one observes for all samples a steep rise of the curves following the solid black line which was added as a guide to the eye. The line indicates the theoretical elongation of the silica clusters assuming a pure affine deformation without any relaxation effects, i.e. linear with strain. The increase of $R_g$ for the clusters in the $\parallel$ direction affinely follows the elongation of the rubber up to about roughly 50\%. Until this strain level, the clusters have a certain degree of freedom to be deformed affinely due to rearrangements of the silica particles that follow the applied strain direction within the cluster.

Beyond 50 \% strain the size changes are less pronounced with strain and a sub-affine deformation regime of the clusters can be observed. 
Most likely, the observed change in $R_g$ growth is caused either by the relaxation of local stresses due to rearrangements and breakage of silica structures or due to the formation of early-stage cavities that are still too small to generate a refraction streak.

With further increasing strain, the $R_g$ of the clusters continues to increase linearly again, however, with different rates according to the composition of the samples.
With decreasing filler levels from sample A to C, the slope formed by the radius of Gyration is continuously increasing since there is the most potential for rearrangements and breakage in highly filled compounds, and hence, more stress relaxation.

Beyond a strain of 200\%, the radius of the clusters does not increase anymore for all three samples thus indicating that the overall forces acting on the clusters exceed the yield strength of the interfacial bonds, which leads to a pronounced local breakage within the inter-phase (cavitation) rather than breakage of the intra-silica bonds of the cluster.

Consequently, beyond this critical macroscopic strain, there is no resulting deformation of silica clusters possible, which is in good agreement with the observed evolution of the $R_g$. Furthermore, $R_g$ is smallest for the highly filled sample A and largest for the lowest filled sample C, which is best explained with a local strain amplification due to the presence of filler and hence a rather early critical strain in a highly filled sample and a more stable interphase in a low filled compound.

The large achievable radius for Sample C also confirms the high intrinsic toughness of the silica clusters which is confirmed by the measurements performed at Continental AG (Table~\ref{addons}) indicating that Sample C is the stiffest.

We can assume that there exists an optimal polymer-to-filler ratio that is beneficial for stronger intra- and inter-cluster interactions resulting in a slower fracture rate of clusters. 
For Samples A and B, a disproportional amount of filler volume fraction as compared to polymer explains the lack of polymer connectivity between silica and polymer. This leads to all stress being transferred to the filler instead of balancing between filler and polymer, resulting in filler jamming and a higher cavitation rate. Therefore, the radius of Gyration for Samples A and B does not reach the same level as for Sample C.

In perpendicular $\perp$ direction, an affine deformation leads to a decrease of sizes according to $R_g\propto E^{-1/2}$ (with $E=1+\epsilon$ being the extension ratio and $E$ being the strain), which is in good agreement with the observed decrease for all samples. Beyond strains close to 100\%, there is no change in observable size most likely due to local stress relaxation effects that lead to sub-affine deformations of the clusters in $\parallel$ direction.

Extending the analysis to lower \textit{q} values between 0.004 nm\textsuperscript{-1} and 0.03 nm\textsuperscript{-1}, we observe a broad bump already well visible in the unstrained samples. It can be fitted with a third structural level in the USF. The fit parameters for the characteristic size in $\parallel$ and $\perp$ directions are shown in Figure \ref{sizes-Rg3}.

For samples A and B, we observe a continuous decrease of this fit parameter in the $\parallel$ direction from about 150 nm to 70 nm, whereas in Sample C it remains constant around 150 nm and rises at strains above 250 \% to $R_g$ more than 200 nm.

We cannot unambiguously determine the origin of the structural parameter represented by the observed broad bump. However, we have reason to think that it represents another hierarchical silica structure, i.e. aggregated silica clusters, rather than a distance between ordered silica clusters. In the latter case, the characteristic distance between more or less densely filled regions in the polymer matrix would be very close to the cluster size and the fact that it decreases with increasing strain would be difficult to explain. 

Interpreting this third structural level as a second order of silica cluster aggregates, formed from mass-fractal silica clusters, can give us further insight into the structural stability of the investigated rubber compound where the polymer-to-filler ratio seems to play an important role. 

For samples A and B, the radius of gyration of these cluster aggregates is rather constant up to 100 \% strain. Afterwards, it decreases until an applied strain of 300 \% which is interpreted as a breakdown into smaller sub-units. For sample C, it remains constant until about 250 \% of strain before the aggregates even expand. As seen before in  Fig. \ref{sizes-Rg2}, the size of the silica clusters increases at the same time continuously until 200 \% of strain.

One has to keep in mind that for samples A and B the share of polymer is less than 50 vol.\%. Hence the majority of the sample is filled with silica or oil which strongly decreases the overall connectivity within the polymer phase and most stresses have to be carried by the various structures forming the filler network. The large-scale filler structures are however not very stable under higher stresses and break down. \textsuperscript{\cite{vilgis_reinforcement_nodate, diani_review_2009}}

In contrast, sample C consists of more than 50 vol.\% of polymer and therefore shows an overall higher connectivity of the polymer network. Better connectivity helps support filler at higher deformations within the polymer matrix resulting in an expansion of the cluster aggregates when the expansion of the silica clusters has reached its maximum size (Fig. \ref{sizes-Rg2}). Eventually, at very high strains these stable filler structures of aggregated clusters also break down due to high local stresses, and hence one observes a decrease in size.

In the $\perp$ direction, the $R_g$ for all three sample types is independent of the applied strain within the error bars of the fit. This aligns well with the assumption of the incompressibility of rubber. \textsuperscript{\cite{benoit_rubber_1988,rivlin_large_1997}}

\section{Conclusions}\label{sec5}

For the first time, we report a systematic study of cavity formation in rubber during \textit{in situ} stretching in the USAXS regime. The increased sample-to-detector distance of 31 m allows us to surpass the limitations observed in previous SAXS studies by probing lower \textit{q} values, thus facilitating the early detection of very weak refraction streaks only visible very close to the direct beam. We describe a methodology to detect the earliest stages of cavitation by identifying the characteristic intensity streak occurring in the direction perpendicular to the applied strain and analyzing the angular dependence of the scattering intensity as a function of the azimuthal angle. Simultaneously from the same datasets, we extract and describe structural information about the hierarchical structure of the silica filler particles within the rubber compound.

To gain statistically robust data, we developed and utilized a sample environment capable of performing high-throughput measurements under quasi-static deformation up to the breaking of the sample.

The newly introduced methodology allowed us to characterize in multiple repetitions the first set of industrial-grade rubber samples with systematically varying production parameters. Samples, where the filler and oil content predominates over the polymer content (Samples A and B), showed higher cavitation rates and instability within filler structures, indicating that the silica filler network is the main recipient of the incoming stresses. On the other hand, the sample, where polymer content is dominant over filler and oil content (Sample C), showed better stability of the filler network due to the overall higher connectivity of interconnected polymer-filler networks resulting in better inter-phase interactions between hierarchical silica structures and the polymer.

The described approach will allow us to perform research on more systematic compound variation, leading to a more profound understanding of material physics related to structural changes caused by adjustments in compound ingredients and applied strain. The introduced methodology simplifies the analysis of damage evolution and aids in enhancing industrial rubber compounds.

To improve the level of confidence regarding the discussed silica-structure mechanics during deformation, more systematic studies using e.g. a wider range of silica concentration has to be performed. This will help identify more precisely the critical concentrations of polymer and filler in a filled rubber nano-composite.

\bmsection*{Author contributions}

Ilya Yakovlev: Conceptualization; Investigation; Visualization; Data Curation; Formal Analysis; Methodology; Writing – Original Draft Preparation.  
Michael Sztucki: Conceptualization; Investigation; Supervision; Formal Analysis; Methodology; Writing – Original Draft Preparation.  
Hossein Ali Karimi-Varzaneh: Conceptualization; Methodology; Resources; Supervision; Validation; Writing – Review \& Editing.  
Jorge Lacayo-Pineda: Conceptualization; Validation; Resources; Supervision; Writing – Review \& Editing.  
Frank Fleck: Conceptualization; methodology; Validation; Writing – Original Draft Preparation.  
Christoph Vatterott: Conceptualization; Resources; Validation.  
Ulrich Giese: Supervision; Validation; Writing – Review \& Editing. 

\bmsection*{Acknowledgments}
We want to express our gratitude to Continental AG for providing in-kind support, training, data, and samples; Theyencheri Narayanan, Manfred Burghammer, and all beamline staff for support during development of the sample environment, experiments, and data analysis; ESRF for provision of beamtime on beamlines ID02 and ID13; Deutsches Institut fur Kautschuktechnologie for providing theoretical and practical training on rubber material science; the InnovaXN for the support of the PhD project.

\bmsection*{Financial disclosure}

None reported.

\bmsection*{Conflict of interest}

The authors declare no potential conflict of interest.

\bibliography{bibliography}


\end{document}